\documentclass[epj-spec]{svjour}
\usepackage{graphicx}
\usepackage{bm}
\begin{document}

\title{Symmetry breaking and clustering in a vibrated granular gas with several macroscopically connected
compartments}
\author{J. Javier Brey \fnmsep\thanks{\email{brey@us.es}} \and R.
Garc\'{\i}a-Rojo \and F. Moreno \and M.J. Ruiz-Montero}

\institute{F\'{\i}sica Te\'{o}rica. Facultad de Fisica. Universidad
de Sevilla. Apartado de Correos 1065. 41080 Sevilla, Spain}

\abstract{ The spontaneous symmetry breaking in a vibro-fluidized
low-density granular gas in three connected compartments is
investigated. When the total number of particles in the system
becomes large enough, particles distribute themselves unequally
among the three compartments. Particles tend to concentrate in one
of the compartments, the other two having the (relatively small)
same average number of particles. A hydrodynamical model that
accurately predicts the bifurcation diagram of the system is
presented. The theory can be easily extended to the case of an
arbitrary number of connected compartments.}

\maketitle

\section{Introduction}
\label{s1} A granular system is an assembly of macroscopic particles
or grains interacting inelastically, i.e. mechanical energy is not
conserved. These systems exhibit many peculiar behaviors as compared
with molecular, elastic systems \cite{JNyB96}. One of them is the
tendency to spontaneously segregate into high and low density
regions, i.e. to form density clusters
\cite{GyZ93,GTyZ93,McyY94,LCyG99,OyU98}. Therefore, they provide
prime examples of formation of structures in far from equilibrium
systems. The physical origin of this effect is the inelastic
character of the interactions between particles. The simplest form
of clustering is presented by freely evolving granular systems, as a
consequence of a long-wavelength hydrodynamic instability
\cite{GyZ93,GTyZ93,McyY94}. However this is an ideal theoretical
state that can not be generated experimentally.

Another interesting situation in which clustering effects show up,
was observed several years ago in a seminal experiment \cite{SyN96}.
A vertically vibrated system of grains was confined in a box
separated into two equal connected parts by a wall of a certain
height. For strong vibration, the particles distribute themselves on
the average equally in the two compartments. Nevertheless, lowering
the intensity of vibration (or the frequency) below a critical
value, the spatial symmetry of the system is spontaneously broken,
in the sense that a steady state is reached in which there is no
equipartition of grains between the two compartments. Moreover,
grains in the compartment with less number of particles have larger
average kinetic energy than those in the high density part. By
considering the exchange of particles between the two compartments
as an effusion process, Eggers \cite{Eg99} proposed an analytical
model to explain this phenomenon. Later on, the original experiment
and also Egger's theory have been extended in different ways,
considering systems with several compartments in which both
clustering and declustering may occur \cite{vMWyL02}, and also
disperse mixtures of grains \cite{MvMWyL02}.

On the other hand, another symmetry breaking mechanism was reported
in \cite{BMGyR01}. In this case, the equilibrium between the two
compartments is hydrodynamic and not merely effusive. Consequently,
this theory applies when the size of the opening connecting both
sides is large as compared with the mean free path of the granular
gas in its neighborhood. This is the opposite limit of that for
which a  balance of the flux of particles is enough to guarantee
stationarity. The extension to  binary mixtures of this hydrodynamic
equilibrium has been investigated in \cite{ByT03}, by means of
molecular dynamics (MD). Let us also mention than while in Eggers'
approach the external gravitations field plays an essential role,
the mechanism put forward in ref. \cite{BMGyR01} is formulated in
the zero gravitational field limit.

The aim of this contribution is to extend the study carried out in
\cite{BMGyR01} to the case of several, more than two, connected
compartments. The size of the holes connecting the compartments is
considered `macroscopic', i. e. larger than the mean free path of
the granular fluid next to it.

It is a pleasure to dedicate this work to our good friend and
colleague Carlos P\'{e}rez. We know he would have enjoyed discussing
this kind of structure formation and had apportted interesting
suggestions as well as key ideas for its better understanding.

\section{The system and molecular dynamics simulation results}
\label{s2} The granular fluid is often modeled as a system of $N$
smooth inelastic hard spheres ($d=3$)  or disks ($d=2$) of mass $m$
and diameter $\sigma $. The inelasticity of the collisions is
described by means of a constant, velocity independent, coefficient
of normal restitution $\alpha$, defined in the interval $0 < \alpha
\leq 1$. This rather simplified  model will be the one considered
here. It has proven to qualitatively capture many of the
experimental features of real granular systems. Then, when two
particles $i$ and $j$ collide having pre-collisional velocities
${\bm v}_{i}$ and ${\bm v}_{j}$, respectively, their
post-collisional  velocities, ${\bm v}^{\prime}_{i}$ and ${\bm
v}^{\prime}_{j}$, are given by
\begin{equation}
\label{2.1} {\bm v}^{\prime}_{i}= {\bm v}_{i} -\frac{1 +\alpha}{2} (
\widehat{\bm \sigma} \cdot {\bm v}_{ij}) \widehat{\bm \sigma},
\end{equation}
\begin{equation}
\label{2.2} {\bm v}^{\prime}_{j}= {\bm v}_{j} +\frac{1 +\alpha}{2} (
\widehat{\bm \sigma} \cdot {\bm v}_{ij}) \widehat{\bm \sigma},
\end{equation}
where ${\bm v}_{ij} \equiv {\bm v}_{i}-{\bm v}_{j}$ is the relative
velocity before collision, and $\widehat{\bm \sigma}$ is the unit
vector directed along the the line joining the centers of the two
particles at contact, away from particle $j$.

\begin{figure}
\begin{center}
\includegraphics[angle=90,scale=0.4]{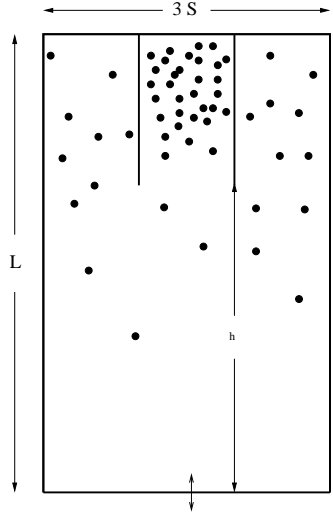}
\end{center}
\caption{Schematic picture of the set up considered. The wall at the
bottom is vibrated in a sawtooth way with velocity $v_{b}$ and small
amplitude, while all the other walls are at rest. Collisions of the
particles with all the walls are elastic} \label{fig1}
\end{figure}

The grains are enclosed in a box of width $3S$ and height $L$. For
$d=2$, $S$ is a length, while for $d=3$, it is an area. The box is
divided into three equal compartments by two walls starting at a
height $h$. A sketch of the system is given  in Fig.\ \ref{fig1}.
Collisions of particles with all the walls are elastic. To keep the
system in a fluidized state, the wall located at the bottom is
vibrated in a sawtooth way with a velocity $v_{b}$ \cite{McyB97}.
This means that all particles colliding with the wall find it moving
'upwards' with that velocity. Moreover, the amplitude of the
vibration is considered much smaller than the mean free path of the
gas in the vicinity of the wall, so the position of the wall can be
taken as fixed. No external field acting on the particles is
considered, something that can be understood as corresponding to the
very strong shaking limit.

\begin{figure}
\begin{center}
\includegraphics[scale=0.3]{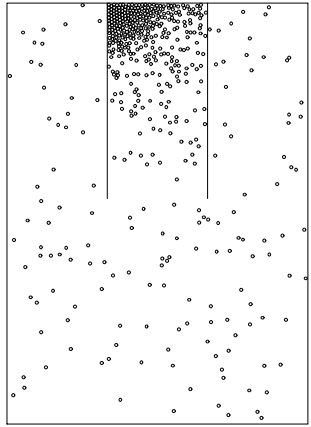}
\end{center}
\caption{Instantaneous snapshot of a typical configuration of a
system in the steady state with broken spatial symmetry. The values
of the parameters are $N=500$, $\alpha=0.9$, $L=140 \sigma$, $S=100
\sigma /3$, $h=75 \sigma$, and $v_{b}=2 \sqrt{ \frac{2 T(0)}{m}}$}
\label{fig2}
\end{figure}

We have performed a series of MD simulations of the system described
above in two dimensions, i.e. for a system of hard disks. In all
cases, the simulations started with the same number of particles in
each of the compartments and a Gaussian velocity distribution. For
given values of $\alpha $ and $v_{b}$, a steady state is reached in
which, on the average, the particles are equally divided into the
three compartments, as long as its total number $N$ is small.
Nevertheless, when $N$ is increased beyond a certain critical value,
the spatial symmetry of the steady state is spontaneously broken.
More specifically, in all the simulations we have carried out, it is
observed that, after a transient period, the particles concentrate
in one of the compartments, the other two having roughly the same
much smaller number of particles. Moreover, particles in the high
density compartment have significantly smaller velocity than
particles in the low density ones. An example of a typical
instantaneous snapshot is given in  Fig.\ \ref{fig2}, which
corresponds to a system of 500 inelastic hard disks with
$\alpha=0.9$. The height of the system is $L=140 \sigma$, and the
width of each of the compartments $S=100 \sigma / 3$. It is clearly
seen that the number of particles in the central compartment is much
larger than in the other two. The average density of the system is
quite low. In particular, a dilute gas theory can be expected to
apply in the lower part of the system, where there is no wall
separating both compartments.

\begin{figure}
\begin{center}
\includegraphics[angle=-90,scale=0.4]{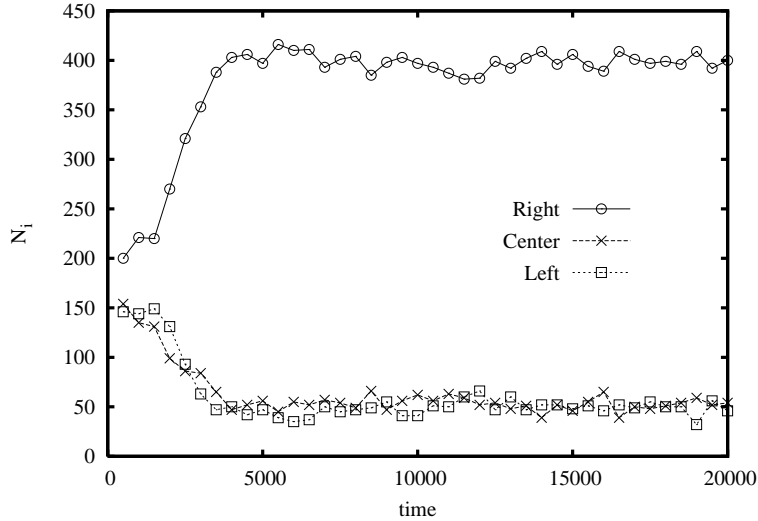}
\end{center}
\caption{Time evolution of the number of particles in each of the
three compartments for the same system as considered in Fig.
\protect{\ref{fig2}}. Time is scaled with the inverse of the
frequency given in Eq. (\protect{\ref{2.3}})} \label{fig3}
\end{figure}

To have a clearer idea of what actually happens in the system, the
time evolution along a trajectory of the population of each of the
compartments for the same system as considered in Fig. \ref{fig2} is
plotted in Fig.\ \ref{fig3}. Time has been scaled with the inverse
of the initial Boltzmann collision frequency $\nu_{0}(0)$ given by
\begin{equation}
\label{2.3} \nu_{0}(0)= \frac{2 N \sigma}{3 S L} \left(
\frac{T(0)}{m} \right)^{1/2},
\end{equation}
where $T(0)$ is the (arbitrary) initial granular temperature of the
particles. As usual, the granular temperature is defined from the
average kinetic energy with the Boltzmann constant set equal to
unity. It is observed that the symmetry of the system is broken very
fast, and the system quickly evolves to a steady state in which the
central compartment has about $400$ particles, while the other other
two compartments have about $50$ each. It is worth to mention that
in the figures shown here in relation with the time evolution of the
system, only a relevant part of the latter is shown. The times
reached in the simulations have always been much larger than those
reported here in the figures, namely up to be really sure that the
system was actually in a steady situation.

\begin{figure}
\begin{center}
\includegraphics[angle=-90,scale=0.4]{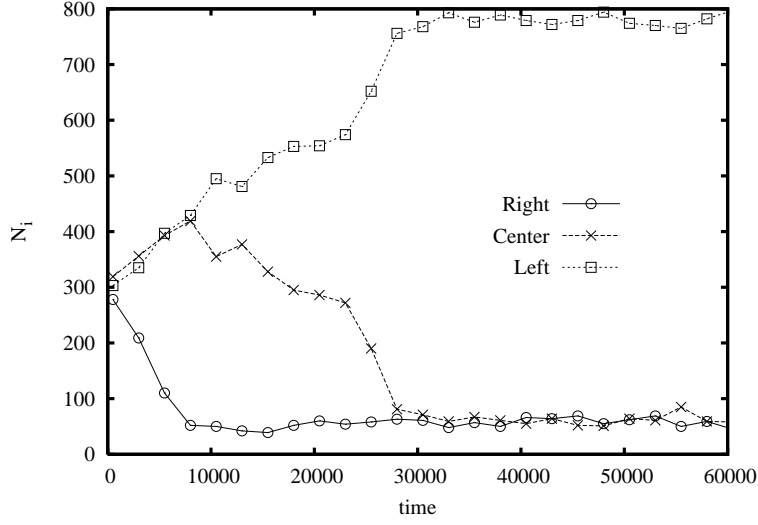}
\end{center}
\caption{Time evolution of the number of particles in each of the
compartments for a system of $N=900$ particles. The values of all
the other parameters are the same as in Fig.\ \protect{\ref{fig2}}.
Note the slow relaxation of the system towards the steady state in
this case as compared with that shown in Fig.\ \protect{\ref{fig3}}}
\label{fig4}
\end{figure}

As already mentioned, in all the carried out simulations, it has
been found that when the spatial symmetry is broken, the particles
agglomerate in one of the compartments, the other two having
approximately the same (low) number of particles. Nevertheless, in
some cases in which the symmetry breaking proceeds rather slowly,
the final state is reached after going through an intermediate
time-dependent unstable state. For a given coefficient of
restitution, this usually happens when the number of particles is
high enough. In the intermediate state, two compartments have the
same  relatively large number of particles, while the number of
particles in the third one is significantly lower. After some time,
this configuration decays, and the number of particles in one of the
populated compartments decreases until reaching the level of the
other low density compartment. An example of this kind of behavior
is provided in Fig.\ \ref{fig4}. The population of the central
compartment grows for a while above the average, having the same
value as that of the left compartment. Afterwards, the number of
particles in it decreases until reaching the same value as in the
right compartment.

\begin{figure}
\begin{center}
\includegraphics[angle=-90,scale=0.4]{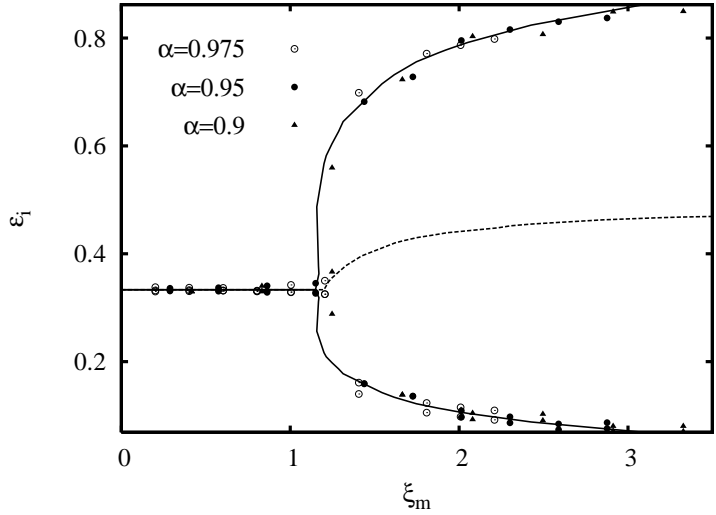}
\end{center}
\caption{Bifurcation diagram of the vibrated system with three
compartments sketched in Fig. \protect{\ref{fig1}}, showing the
relative average number of particles in each compartment
$\epsilon_{i} \equiv \overline{N}_{i}/N $, as a function of the
dimensionless parameter $\xi_{m}$ defined in the text. The symbols
are MD results for the values of $\alpha$ indicated. The other
parameters of the system are: $L=140 \sigma$, $S= 100 \sigma /3$,
and $v_{b}= 2 \sqrt{2 T(0)/m}$. The lines are the theoretical
prediction from the model developed in the text, where its meaning
is explained} \label{fig5}
\end{figure}

The symmetry or asymmetry of the steady state can be characterized
by the set of parameters
\begin{equation}
\label{2.4} \epsilon_{i} \equiv  \frac{\overline{N}_{i}}{N},
\end{equation}
with $\overline{N}_{i}$ being the average number of particles in the
compartment $i$ in the steady state. Of course, in a symmetric
configuration, it is $\epsilon_{i}=1/3$ for all $i$. The resulting
bifurcation diagram is shown in Fig. \ref{fig5}, where the three
parameters $\epsilon_{i}$ are plotted as a function of a
dimensionless control parameter $\xi_{m}$, that is proportional to
the total number of particles and will be defined below. The data
for different values of $\alpha$ are seen to collapse on the same
curves. Moreover, it has been found that the bifurcation diagram is
not altered by modifying the velocity of the wall $v_{b}$, as long
as it is large enough as to keep all the granular system fluidized.
For small values of $\xi_{m}$, the three compartments are equally
populated, but for $\xi_{m} > \xi_{0}$, one of the compartments has
a larger number of particles than the other two. Note that the same
symbol is used in the figure for the three compartments since they
are equivalent, in the sense that they interchange their populations
in different trajectories of the same system.

\section{The model}
\label{s3} Consider the steady state reached by a vibrated dilute
granular gas with only one compartment of width $3S$ and assume
there are gradients only in the vertical direction, taken as the $x$
axis. It is convenient to define a dimensionless space scale by
\begin{equation}
\label{3.1} \xi = \sqrt{a(\alpha)} \int_{0}^{x} \frac{d
x^{\prime}}{\lambda (x^{\prime})}\, ,
\end{equation}
where $\lambda (x)$ is the local mean free path given by
\begin{equation}
\label{3.2} \lambda (x)= \left[ C_{d} \sigma^{d-1} n(x)
\right]^{-1},
\end{equation}
with $C_{2}=2 \sqrt{2}$, $C_{3}= \pi \sqrt{2}$, $n(x)$ the local
number density, and
\begin{equation}
\label{3.3} a(\alpha)=\frac{32(d-1)\pi^{d-1}
\zeta^{*}(\alpha)}{(d+2)^{3}C_{d}^{2}\Gamma^{2} \left(d/2 \right)
[\kappa^{*}(\alpha)-\mu^{*}(\alpha)]}\, .
\end{equation}
Here $\kappa^{*}(\alpha)$ and $\mu^{*}(\alpha)$ are the
dimensionless transport coefficients characterizing  the
Navier-Stokes heat flux in a granular gas \cite{BDKyS98}. Finally,
$\zeta^{*}(\alpha)$ is the dimensionless cooling rate due to the
energy dissipation in collisions. The explicit expressions of these
quantities can be found in ref. \cite{BRyM00}. For elastic systems
with  $\alpha \rightarrow 1$, $\kappa^{*}$ tends to unity, while
$\mu^{*}$ and $\zeta^{*}$ vanish.

\begin{figure}
\begin{center}
\includegraphics[angle=-90,scale=0.35]{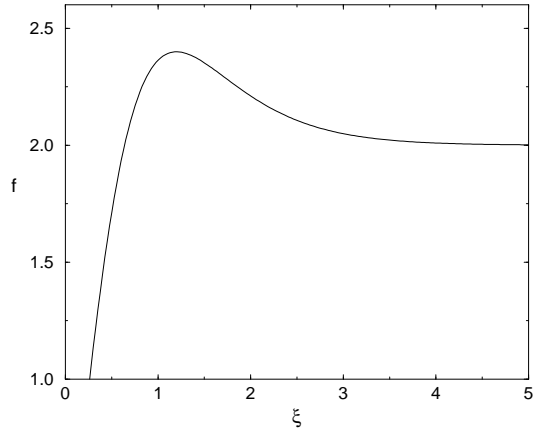}
\end{center}
\caption{The function $f(\xi)$ defined in Eq.\
(\protect{\ref{3.5}}). It characterizes the dependence of the
pressure on the total number of particles in the one-compartment
system} \label{fig6}
\end{figure}

In the upper limit of the system $x=L$, $\xi$ takes its maximum
value $\xi_{m} \equiv  \xi(x=L)= \sqrt{a(\alpha)} C_{d} \sigma^{d-1}
N_{x}$, with $N_{x} \equiv N/3S$ being the number of particles per
unit of section of the vibrating wall. This is the quantity used in
the horizontal axis of Fig.\ \ref{fig5}. In the Navier-Stokes
approximation, the hydrodynamic pressure of the system is uniform
and can be expressed as \cite{BRyM00}
\begin{equation}
\label{3.4} p= \frac{T_{0}}{4 C_{d} \sigma^{d-1} L \sqrt{a(\alpha)}}
f(\xi_{m}),
\end{equation}
where $T_{0}$ is the temperature of the gas next to the vibrating
wall and the function
\begin{equation}
\label{3.5} f(\xi) \equiv \frac{2 \xi+\sinh (2 \xi)}{\cosh^{2}
\xi}\,
\end{equation}
has been introduced. This function is plotted in Fig. \ref{fig6}. It
presents a maximum at $\xi=\xi_{0}$, where $\xi_{0} \simeq 1.20$ is
the non-zero solution of the equation $\xi_{0} \tanh \xi_{0} =1$.
For large values of $\xi$ it tends asymptotically to $2$.

Now come back to the steady state reached by the system with the
three compartments represented in Fig.\ \ref{fig1}. By extending the
ideas developed in \cite{BMGyR01}, we model this state by treating
the compartments as independent, sharing only a thin, but
macroscopic, layer of granular gas next to the vibrating wall at the
temperature $T_{0}$. Macroscopic, hydrodynamic equilibrium between
the three compartments requires that the pressure be the same in all
them, and use of Eq. (\ref{3.4}) gives
\begin{equation}
\label{3.6} f(\xi_{m}^{(1)})  = f(\xi_{m}^{(2)}) = f(\xi_{m}^{(3)}),
\end{equation}
where $\xi_{m}^{(i)}$ is the maximum value of $\xi$ in the
compartment $i$, i.e. next to the upper wall,
\begin{equation}
\label{3.7} \xi_{m}^{(i)} \equiv  \sqrt{a(\alpha)} C_{d}
\sigma^{d-1} \frac{\overline{N}_{i}}{S}.
\end{equation}
Since the total number of particles in the three compartments is
$N$, the quantities $\xi_{m}^{(i)}$ must verify the condition
\begin{equation}
\label{3.8} \xi_{m}^{(1)}+ \xi_{m}^{(2)}+\xi_{m}^{(3)} = 3 \xi_{m}=
3 \sqrt{a(\alpha)} C_{d} \sigma^{d-1} N_{x}.
\end{equation}
Of course, Eqs.\ (\ref{3.6}) always have the symmetric solution
$\xi_{m}^{(1)}=\xi_{m}^{(2)}=\xi_{m}^{(3)}= \xi_{m}$, in which the
three compartments have the same average number of particles. From
the observation of Fig.\ \ref{fig6}, it follows that this is the
only solution for $\xi_{m} < \xi_{0} \simeq 1.20$. On the other
hand, for $\xi_{m} > \xi_{0}$ other solutions are possible since
there are two different values $\xi^{-}$ and $\xi^{+}$, $\xi^{-} <
\xi^{+}$, for which $f(\xi^{-})=f(\xi^{+})$. Therefore, in this
parameter region two asymmetric solutions are possible:
\begin{itemize}
\item {\em Asymmetric steady state I}. Two compartments have the same average number
of particles, that is {\em smaller} than the average number of
particles in the third one. Mathematically, it is defined by
\begin{equation}
\label{3.9}  2 \xi_{m}^{-(I)} + \xi_{m}^{+(I)} = \xi_{m}.
\end{equation}

\item {\em Asymmetric steady state II}. Two compartments have the same
average number of particles, that is {\em larger} than the average
number of particles in the third one. It is defined by
\begin{equation}
\label{3.10}  \xi_{m}^{-(II)} + 2 \xi_{m}^{+(II)} = \xi_{m}.
\end{equation}

\end{itemize}
Given the value of $\xi_{m} > \xi_{0}$ characterizing the system
under consideration, the two above asymmetric solutions are found by
numerically solving the equation $f(\xi^{-})=f(\xi^{+})$ together
with Eq.\ (\ref{3.9}) or (\ref{3.10}), respectively. This solutions
are the lines plotted in Fig.\ \ref{fig5}. The upper (lower) solid
line is the function $ \xi_{m}^{+(I)}$ ($ \xi_{m}^{+(II)}$). The
dashed line is the function $\xi_{m}^{+(II)}$. The function
$\xi_{m}^{-(II)}$ has not been plotted for the sake of clearness,
taken into account that in the simulations, the system never was
found in the asymmetric steady state II. Therefore, it seems that
this state is unstable. It is seen that the agreement between the
predicted values of the relative populations in the state $II$ and
the simulation results is pretty good. In particular, it is observed
that their dependence on the coefficient of restitution $\alpha$ is
scaled out when $\xi_{m}$ is used as the control parameter.

The asymmetry of the stable state $I$ increases very fast as the
total number of particles in the system increases. Even more, it can
be observed in the figure that the steady average number of
particles in the two less populated compartments decreases as more
particles are added to the system. Let us stress that, even when the
number of particles in these compartments is rather small, the
hydrodynamic description presented here describes quite accurately
the asymmetry of the system.

\section{Conclusions}
\label{s4} It has been shown that the problem of the spontaneous
symmetry breaking in a vibrated granular gas with three compartments
connected by a macroscopic hole, can be treated in a way similar to
the case of two compartments. In fact, the theory can be easily
extended to deal with an arbitrary number of compartments. The only
important point to be verified is that the hydrodynamic pressure and
temperature near the vibrating wall are the same in all the
compartments. In the simulations presented here, it was found that
they agree within the numerical precision of the measurements. Of
course, the exhibited symmetry breaking becomes much richer as the
number of compartments is increased.

An important point not yet well understood is why the system
actually chooses the asymmetric state $I$  with preference to the
symmetric state and to the asymmetric state $II$, i.e. why the
latter two are unstable. In principle, relevant information might be
obtained by carrying out an stability analysis of the solutions.
Nevertheless, such analysis is rather involved and has not been
completed up to now. On the other hand, it is possible to compute
the power dissipated in collisions for each of the states. This is
done by taking into account that the energy balance requires that,
in the steady state, the dissipated power be the same as the one
injected through the vibrating wall. The latter for a system with a
compartment is given by $v_{b} S p$. For a system with three
compartments, the powers dissipated in each of them have to be added
up. Then, a simple analysis shows that it is maximum for the
asymmetric state $I$. Also in the system with two compartments the
asymmetric state dissipates more energy than the symmetric one.

It is worth to mention that some care is needed when studying the
state discussed in this paper both by numerical simulations and also
by experiments. The steady state of the system with one compartment
discussed at the beginning of Sec. \ref{s3} is known to be unstable
when its width is larger than a certain critical value. More
precisely, the system experiments a continuos spontaneous symmetry
breaking in the direction perpendicular to the heat flux, next to
the  wall opposite to the vibrating one \cite{LMyS02,BRMyG02}. The
existence of this instability must be taken into account when
deciding the sizes of the system to be used.

\begin{acknowledgement}
This research was supported by the Ministerio de Educaci\'{o}n y
Cienc\'{\i}a (Spain) through Grant No. FIS2005-01398 (partially
financed by FEDER funds).
\end{acknowledgement}

\end{document}